\documentclass[release,onefignum,onetabnum]{siamart171218}
\pdfoutput=1 

\usepackage{lipsum}
\usepackage{amsfonts}
\usepackage{graphicx}
\usepackage{subcaption}
\usepackage{epstopdf}
\usepackage{algorithmic}
\ifpdf
  \DeclareGraphicsExtensions{.eps,.pdf,.png,.jpg}
\else
  \DeclareGraphicsExtensions{.eps}
\fi

\newsiamremark{remark}{Remark}
\newsiamremark{hypothesis}{Hypothesis}
\crefname{hypothesis}{Hypothesis}{Hypotheses}
\newsiamthm{claim}{Claim}

\headers{Acceleration of the Power Method with DMD}{J. Roberts, L. Xu, R. Elzohery, M. Abdo}

\title{Acceleration of the Power Method with Dynamic Mode Decomposition}

\author{Jeremy A. Roberts\thanks{Department of Mechanical and Nuclear Engineering, Kansas State University, Manhattan, KS 66506
  (\email{jaroberts@ksu.edu})} \and Leidong Xu \and Rabab Elzohery \and Mohammad Abdo}

\usepackage{amsopn}

\ifpdf
\hypersetup{
  pdftitle={Acceleration of the Power Method with Dynamic Mode Decomposition},
  pdfauthor={J.A. Roberts, L. Xu, R. Elzohery, M. Abdo}
}
\fi

\usepackage{bm}

\begin{document}

\maketitle

\begin{abstract}
Presented is an algorithm based on dynamic mode decomposition (DMD) for acceleration of the power method (PM).
The power method is a simple technique for determining the dominant eigenmode of an operator $\mathbf{A}$, and variants of the power method are widely used in reactor analysis.
Dynamic mode decomposition is an algorithm for decomposing a time-series of spatially-dependent data and producing an explicit-in-time reconstruction for that data.
By viewing successive power-method iterates as snapshots of a time-varying system tending toward a steady state, DMD can be used to predict that steady state using (a sometime surprisingly small) $n$ iterates.
The process of generating snapshots with the power method and extrapolating forward with DMD can be repeated.
The resulting restarted, DMD-accelerated power method (or DMD-PM($n$)) was applied to the two-dimensional IAEA diffusion benchmark and compared to the unaccelerated power method and Arnoldi's method.
Results indicate that DMD-PM($n$) can reduce the number of power iterations required by approximately a factor of 5.
However, Arnoldi's method always outperformed DMD-PM($n$) for an equivalent number of matrix-vector products $\mathbf{Av}$.
In other words, DMD-PM($n$) cannot compete with leading eigensolvers if one is not limited to snapshots produced by the power method.
Contrarily, DMD-PM($n$) can be readily applied as a post process to existing power-method applications for which Arnoldi and similar methods are not directly applicable.
A slight variation of the method was also found to produce reasonable approximations to the first and second harmonics without substantially affecting convergence of the dominant mode. 
\end{abstract}



\section{Introduction}

Eigenvalue problems often arise from equations that describe steady-state systems.
In some applications, the fundamental (or dominant) eigenmode (i.e., the eigenvector corresponding to the eigenvalue of largest magnitude) represents the state of the system that would be observed in true, steady-state conditions.
One such application arises in the analysis of nuclear reactors, for which the balance of neutrons is modeled using a linearized Boltzmann equation or a related diffusion approximation; either can be represented by the generalized eigenvalue problem
\begin{equation}
 \mathbf{F} \mathbf{x} =  k \mathbf{L} \mathbf{x}  \, ,
 \label{eq:keig}
\end{equation}
where $\mathbf{x} $ describes the neutron population, $\mathbf{L}$ represents system losses, $\mathbf{F}$ represents system gains, and the eigenvalue $k = ||\mathbf{F}\mathbf{x} || / ||\mathbf{L}\mathbf{x} ||$ represents the ratio of gains to losses. 

To determine the fundamental mode, a variety of algorithms exist, but perhaps the simplest is the power method.
The application of the power method to a steady-state system (e.g., a nuclear reactor model) can be interpreted as follows.
The system is initialized in an unsteady state corresponding to some linear combination of the system modes (i.e., the eigenvectors of the operator $\mathbf{L}^{-1}\mathbf{F}$).
The dominant mode has a corresponding frequency $\omega_0 = 0$, i.e., its temporal evolution is governed by $e^{\omega_0 t} = 1$.
All higher-order modes have frequencies $\omega_i < 0$ for $i > 0$ and, therefore, decay in time.
Each iteration of the power method moves the system forward one characteristic step in time, and, therefore, the higher-order modes decay at each step.
The method continues until these higher-order modes are sufficiently decayed.

These frequencies and the spatial shapes to which they correspond are precisely what dynamic mode decomposition (DMD) reveals.
First proposed to infer system dynamics from a time series of fluid flow-field observations \cite{schmid:hal-01053394, schmid2010dynamic}, the basic DMD algorithm has been rigorously analyzed mathematically \cite{tu2013dynamic}, and extended to a variety of applications thoroughly summarized in a recent monograph \cite{kutzbook}.  
In some applications, DMD has been used to recover approximations to physical parameters from the output of existing (possibly black-box) models that would otherwise require significant changes be made to those models.  
For example, DMD was used to recover $\alpha$ eigenvalues from time-dependent, neutron transport calculations \cite{mcclarren2019calculating}.  
Others have used DMD as a direct, explicit-in-time surrogate for such black-box models, e.g., to model the evolution of nuclear reactor isotopics over long time periods \cite{abdo2019modeling, elzohery2018cbg}, the nonlinear response of reactor power during short transients \cite{LRA}, and nuclear-fuel, decay-heat generation \cite{alfonsi2018dhc}.

Of particular relevance here are two past applications of DMD to accelerate existing, iterative methods.
In perhaps the first such application, a modified DMD was used to accelerate the convergence of time-dependent, computational fluid dynamics models to their steady state by starting from (non-steady) initial conditions and marching forward in time until converged \cite{andersson2014novel}. 
In that method, a system is integrated over a time step to produce a sequence of flow-field snapshots of  the form $\mathbf{x}^{(n+1)} = \mathbf{A}\mathbf{x}^{(n)} + \mathbf{b}$.  
With use of $\mathbf{X}_{n} = [\mathbf{x}^{(2)}-\mathbf{x}^{(1)}, \ldots, \mathbf{x}^{(n+1)}-\mathbf{x}^{(n)}]$, a low-rank approximation $\tilde{\mathbf{A}}$ can be formed using the basic idea of DMD to be discussed Section~\ref{sec:dmd} and used to approximate the correction $\Delta\mathbf{x}$ that satisfies the steady-state condition $(\mathbf{I}-\mathbf{A})\Delta \mathbf{x} = (\mathbf{A}-\mathbf{I})\mathbf{x}^{(n+1)} + \mathbf{b}$.
The corrected steady-state solution is $\mathbf{x} = \mathbf{x}^{(n+1)}+\Delta\mathbf{x}$.
By performing several rounds of time stepping followed by a correction, both the number of iterations and the average fluctuation in successive iterates was reduced for the problems studied\cite{andersson2014novel}.  

In a similar fashion, DMD was applied to source-driven neutronics problems through acceleration of Richardson (or ``source'') iteration for the inhomogeneous equation $\mathbf{L}\mathbf{x}  = \mathbf{b}$ (where $\mathbf{L}$ can be the same as in Eq.~(\ref{eq:keig})) \cite{mcclarren2018acceleration}.  Richardson iteration leads to the sequence $\mathbf{x}^{(n+1)} = (\mathbf{I}-\mathbf{L})\mathbf{x}^{(n)} + \mathbf{b}$.  As in Ref.~\cite{andersson2014novel}, a set of successive differences $\mathbf{x}^{(n)}-\mathbf{x}^{(n-1)}$ can be used to produce an approximate operator $\tilde{\mathbf{A}}$, and a correction $\Delta \mathbf{x}$ can be used to provide an improved estimate for $\mathbf{x}= \mathbf{x}^{(n)}+ \Delta \mathbf{x}$.  Results from several test cases suggest that a sequence of Richardson iterations followed by corrections reduces the number of iterations required by about one order of magnitude.

Building on these past successes, it is shown here that DMD can be used with the power method to estimate the frequencies and spatial modes of a reactor system as it converges to its fundamental, steady-state mode.
The system can then be projected forward in ``time'' to recover a solution that would be observed after (possibly many) additional iterations of the power method.
This process can be repeated, i.e., the power method can provide snapshots of the system in time, and DMD can be used to extrapolate forward in time based on those snapshots to produce a starting point for an addition series of restarted power-method/DMD iterations.

\section{Methodology}

\subsection{The Power Method}
\label{sec:pi}

The power method is a simple, iterative scheme used to determine the dominant eigenpair of the system

\begin{equation}
\mathbf{Ax} = \lambda \mathbf{x} \, .
\end{equation}
Let $\lambda_i$ represent the $i$th eigenvalue of $\mathbf{A}$.  
Further, let these eigenvalues be numbered in decreasing order based magnitude, i.e., $|\lambda_0| > |\lambda_1| \geq |\lambda_2| \ldots$.  
The dominant eigenpair are those $\lambda_0$ and $\mathbf{x}_0$ that satisfy $\mathbf{Ax}_0 = \lambda_0 \mathbf{x}_0$.

The power method proceeds as follows:
\begin{enumerate}
  \item Guess $\mathbf{x}^{(0)}$ and normalize it such that $||\mathbf{x}^{(0)}||=1$.
  \item Update $\mathbf{x}^{(i)} = \mathbf{A}\mathbf{x}^{(i-1)}$ and $\lambda^{(i)} = ||\mathbf{x}^{(i-1)}||$, where $(i)$ represents the iteration.
  \item Set $\mathbf{x}^{(i)} = \mathbf{x}^{(i)} / \lambda^{(i)}$.
  \item Repeat steps 2 and 3 for $i = 1, 2, \ldots$ until $||\mathbf{x}^{(i)} - \mathbf{x}^{(i-1)}|| < \tau$ for some tolerance $\tau$.
\end{enumerate}
As long as the fundamental mode (and eigenvalue) are real and the initial guess $\mathbf{x}^{(0)}$ is not perpendicular to the fundamental mode $\mathbf{x}_0$ (i.e., $\mathbf{x}^T_0\mathbf{x}^{(0)} \neq 0$), the power method will converge to the dominant eigenpair $(\mathbf{x}_0, \lambda_0)$.

The rate of convergence of the power method depends on the relative magnitudes of the leading eigenvalues.  
Let the initial guess be represented as a weighted sum of the eigenvectors of $\mathbf{A}$, i.e.,
\begin{equation}
\begin{split}
  \mathbf{x}^{(0)} 
     &=  c_0' \mathbf{x}_0 + c_1' \mathbf{x}_1 + c_2'  \mathbf{x}_2 \ldots \,  \\
     &=  c_0' \left (\mathbf{x}_0 + \frac{c_1'}{c_0'} \mathbf{x}_1 + \frac{c_2'}{c_0'} \mathbf{x}_2 \ldots  \right ) \\
     &=  c_0' \left (\mathbf{x}_0 + c_1 \mathbf{x}_1 + c_2 \mathbf{x}_2 \ldots  \right )\, .         
\end{split}
\end{equation}
Because normalization of an eigenvector is arbitrary, let $c_0' = 1$.
Then, multiplication of $\mathbf{A}$ by this initial guess leads to
\begin{equation}
\begin{split}
  \mathbf{A} \mathbf{x}^{(0)} 
   &=  \mathbf{A} \mathbf{x}_0  +  c_1 \mathbf{A} \mathbf{x}_1 + c_2 \mathbf{A} \mathbf{x}_2 + \ldots \\
   &= \lambda_0 \mathbf{x}_0 + c_1 \lambda_1 \mathbf{x}_1 + c_2 \lambda_2 \mathbf{x}_2 + \ldots \\ 
   &= \lambda_0 \left ( \mathbf{x}_0 + c_1 \frac{\lambda_1}{\lambda_0} \mathbf{x}_1 + c_2 \frac{\lambda_2}{\lambda_0} \mathbf{x}_2 + \ldots \right ) \, .
\end{split}
\end{equation}
Repeated application of $\mathbf{A}$ leads to
\begin{equation}
\begin{split}
  \mathbf{A}^n \mathbf{x}^{(0)} 
   &= \lambda^n_0 \left ( \mathbf{x}_0 + c_1 \left( \frac{\lambda_1}{\lambda_0} \right)^n \mathbf{x}_1 + c_2 \left ( \frac{\lambda_2}{\lambda_0} \right )^n \mathbf{x}_2 + \ldots \right ) \, ,
\end{split}
\end{equation}
which shows that if $|\lambda_0| > |\lambda_1|$, then $\mathbf{A}^n \mathbf{x}^{(0)}$ will tend toward the direction $\mathbf{x}_0$ at a rate governed by the dominance ratio $|\lambda_1|/|\lambda_0|$. 
Because $\lambda^n_0$ may grow without bound (or vanish), normalization is required during the iteration as is included in the algorithm above.

\subsection{Dynamic Mode Decomposition}
\label{sec:dmd}

The dynamic mode decomposition has been widely described in the literature, and the concise overview given here adapts the lucid presentation given in Ref.~\cite{kutzbook}.
To start, consider the generic, dynamic problem defined by
\begin{equation}
  \frac{d{\mathbf{x}}}{dt}=\mathbf{f}(\mathbf{x},t) \, ,
  \label{eq:dynamic_problem}
\end{equation}
where $\mathbf{x} \in \mathbb{R}^{n}$ is the $n$-dimensional state vector at time $t$. 
Suppose that the evolution of $\mathbf{x}$ can be well approximated by a relationship of the form  
\begin{equation}
 \frac{d{\mathbf{x}}(t)}{dt}=\mathcal{A}\mathbf{x} \, ,
 \label{eq:linearized_model}
\end{equation}
where the mapping operator $\mathcal{A}$ may not be known explicitly. 
However, if one has a sequence of samples $\mathbf{x}(t_i)$ for $t_i = \Delta i$, $i = 0, 1, \ldots, m$ and defines 
\begin{equation}
\label{eq:past_data}
\mathbf{X_0}=\left[ \mathbf{x}_0, \mathbf{x}_1, \ldots, \mathbf{x}_{m-1} \right] 
  \qquad \text{and} \qquad 
\mathbf{X_1}=\left[ \mathbf{x}_1, \mathbf{x}_2, \ldots, \mathbf{x}_{m} \right] \, ,
\end{equation}
then DMD yields the operator $\mathbf{A}$ (or, as discussed below, lower-rank approximations to it) that best satisfies
\begin{equation}
  \mathbf{X}_1 \approx \mathbf{A}\mathbf{X}_0 \, ,
\end{equation}
in a least-squares sense.
Because the solution to Eq.~(\ref{eq:linearized_model}) is $\mathbf{x}(t) = e^{\mathcal{A}t}\mathbf{x}(0)$, $\mathbf{A}$ is the discrete-time approximation to $e^{\mathcal{A}\Delta}$.
Moreover, the discrete eigenvalues $\lambda_i$ of $\mathbf{A}$ are related to the continuous-time eigenvalues $\omega_i$ of $\mathcal{A}$ by
\begin{equation}
 \omega_i = \frac{\ln(\lambda_i)}{\Delta} \, .
\end{equation}
Corresponding to $\lambda_i$ is the eigenvector $\mathbf{z}_i$ of $\mathbf{A}$, and the complete set of eigenpairs of $\mathbf{A}$ yields
\begin{equation}
 \mathbf{x}(t) \approx \sum^{m}_{i=1} \mathbf{z}_i e^{\omega_i t} b_i = \mathbf{Z} e^{\bm{\Omega}t} \mathbf{b} \, ,
 \label{eq:dmd_surrogate}
\end{equation}
and, to satisfy $\mathbf{x}(0) = \mathbf{Z} \mathbf{b}$ in a least-squares sense, let $\mathbf{b} = \mathbf{Z}^{\dagger} \mathbf{x}(0)$, where $\dagger$ indicates the Moore-Penrose pseudoinverse.

The best-fit operator $\mathbf{A}$ is formally given by
\begin{equation}
 \mathbf{A} = \mathbf{X}_1 \mathbf{X}^{\dagger}_0 \, ,
\end{equation}
where $\mathbf{X}^{\dagger}_0$ can be defined from its (thin) singular-value decomposition, i.e.,
\begin{equation}
  \mathbf{X}_0 = \mathbf{U} \bm{\Sigma} \mathbf{V}^{*} \rightarrow \mathbf{X}_0^{\dagger} = \mathbf{V} \bm{\Sigma}^{-1} \mathbf{U}^* \, ,
\end{equation}
where $\mathbf{U} \in \mathbb{C}^{m\times n}$, $\mathbf{V} \in \mathbb{C}^{n\times n}$, $\bm{\Sigma} \in \mathbb{C}^{n\times n}$, and $*$ indicates the conjugate transpose.  
It follows that $\mathbf{A} = \mathbf{X}_1 \mathbf{V} \bm{\Sigma}^{-1} \mathbf{U}^*$.
However, this matrix is large, and, in practice, the low-rank approximation ${\mathbf{\tilde{A}} = \mathbf{U}^*_r \mathbf{A}\mathbf{U}_r}$ can be used.
Here, $\mathbf{U}_r$ contains the first $r < n$ columns of the left singular matrix, usually arranged to correspond to the largest $r$ singular values (but alternative selection schemes have been explored \cite{kutzbook})
Finally, the leading $r$ eigenvalues of $\mathbf{A}$ are inferred from the eigenvalues of $\mathbf{\tilde{A}}$, while the corresponding eigenvectors $\mathbf{Z}$ of $\mathbf{A}$ are recovered from the eigenvectors $\mathbf{W}$ of $\tilde{\mathbf{A}}$ through $\mathbf{Z} = \mathbf{X}_1 \mathbf{V}_r \bm{\Sigma}_r^{-1} \mathbf{W}$.

\subsection{An Accelerated Power Method using DMD}
\label{sec:dmdpi}

The power method for the system $\mathbf{A}\mathbf{x} =\lambda \mathbf{x}$ can be represented by the sequence 
\begin{equation}
 \mathbf{x}^{(i+1)} = \mathbf{f}(\mathbf{x}^{(i)}) = \frac{\mathbf{A}\mathbf{x}^{(i)}}{||\mathbf{A}\mathbf{x}^{(i)}||} \, , \qquad i = 0, 1, \ldots
 \label{eq:power_iter_seq}
\end{equation}
Suppose that $m$ power iterations have been performed to produce $\mathbf{X}_0$ and $\mathbf{X}_1$, where each iterate is separated by a (fictional) $\Delta$ in time.
Application of DMD leads to a set of approximate eigenvectors for $\mathbf{A}$ and a set of eigenvalues $e_j \approx \lambda_j / \lambda_0$.  
The leading eigenvalue $e_0$ tends toward unity, indicating a stationary process (and the convergence of the power method to the fundamental mode).

The normalization in Eq.~(\ref{eq:power_iter_seq}) may be important for reducing numerical round-off errors introduced by growing (or decaying) iterates.  
However, the normalization introduces nonlinearity to $\mathbf{f}$.
This nonlinearity has not led to numerical difficulties, and this may be explained by the strong connection between DMD and the Koopman operator \cite{kutzbook,budivsic2012applied}, which is the infinite-dimensional, linear operator $\mathcal{K}$ that maps $\mathbf{g}(\mathbf{x}(t_{n+1})) = \mathcal{K}\mathbf{g}(\mathbf{x}(t_n))$ for some (possibly nonlinear) observation function $\mathbf{g}(\mathbf{x})$ and iterates $\mathbf{x}(t_i)$ of Eq.~(\ref{eq:dynamic_problem}).

In order to accelerate the power method with DMD, the following DMD-PM($n$) algorithm is proposed:
\begin{enumerate}
 \item Guess $\mathbf{x}^{(0)}$ and normalize.
 \item Perform $n$ power iterations to produce $\mathbf{X}_0$ and $\mathbf{X}_1$
 \item Compute the DMD modes and frequencies using a rank-$r$, truncated  SVD (i.e., $r < n$)
 \item Apply Eq.~(\ref{eq:dmd_surrogate}) to estimate $\mathbf{x}^{(\infty)}=\mathbf{x}(\infty)$, i.e., estimate the steady-state, dominant mode after an equivalent of $\infty$ power iterations.
 \item Set $\mathbf{x}^{(0)} = \Re(\mathbf{x}(\infty)) / ||\mathbf{x}(\infty)||$.  
 \item Repeat Steps 1 through 5 until converged.
\end{enumerate}
In practice, a large value can be used in place for $\infty$.  
Alternatively, because the power method must reveal a single, dominant mode if the assumptions described in Section~\ref{sec:pi} are satisfied, then only one mode recovered by DMD should remain at $t=\infty$.  
In other words, one can directly set $\mathbf{x}^{(0)} = b_0 \mathbf{z}_0 / ||b_0 \mathbf{z}_0||$ (where $b_0$ ensures a consistent sign).

One numerical challenge faced by the algorithm described is that after several passes, the iterates produced in Step 2 are increasingly ill conditioned.  
Evaluation of the complete (reduced) SVD of rank ($n-1$) may lead to numerical errors, and to mitigate these effects, an optimal rank may be selected \cite{gavish2014optimal}.
In addition, it was found that exclusion of the DMD-corrected iterate generated in Step 4 from $\mathbf{X}_0$ improves the numerical stability; a further study is warranted to understand the impact of the initial vector on DMD's performance.

As described, the DMD-accelerated power method constructs a sequence of $n$ vectors that form a basis for an $n$-dimensional Krylov subspace and attempts to extract spectral properties about $\mathbf{A}$ from that basis.
This process bears obvious similarity to the Arnoldi method, with a major difference being that the DMD acceleration post-processes the existing basis while the Arnoldi method continually refines the basis through orthonormalization.
Based on scoping studies, it would appear that given $n$ possible applications of $\mathbf{A}$ to a starting vector $\mathbf{x}^{(0)}$, the Arnoldi method is almost certain to provide a better estimate for $\mathbf{x}_0$ than DMD-PM.  

\section{Numerical Results}

\begin{figure*}[t!]
 \centering
 \includegraphics[width=0.7\textwidth]{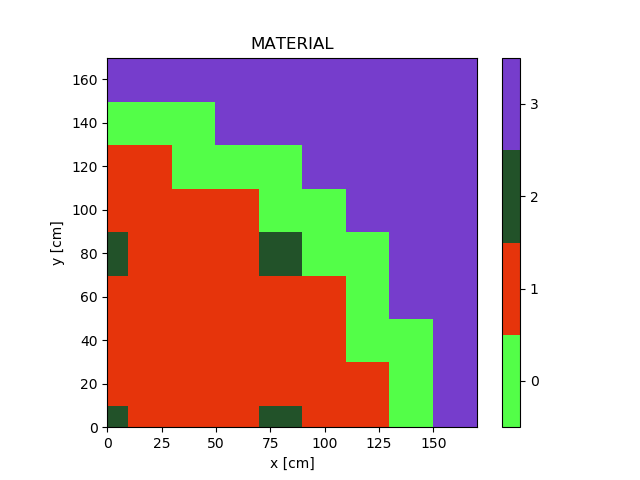}\\
   \caption{Geometry as modeled for the IAEA 2-D diffusion benchmark.  Material properties can be found in the benchmark documentation \cite{center1977benchmark}.  Materials 0 and 1 are fuel, material 2 represents control, while material 3 represents the outer reflector.}
  \label{fig:iaea2d}
\end{figure*}

To illustrate the DMD-PM algorithm proposed in Section~\ref{sec:dmdpi}, it was used to determine the fundamental eigenmode for the well-known, 2-D IAEA diffusion benchmark \cite{center1977benchmark}.  Shown in Figure~\ref{fig:iaea2d} is the basic core layout.
The governing equations are
\begin{equation}
 \begin{split}
  -\nabla D(\mathbf{r})_1 \nabla \phi_1 (\mathbf{r}) + \Sigma_{r1}(\mathbf{r}) \phi_1 (\mathbf{r}) 
    &= \frac{1}{k} \left (\nu\Sigma_{f1}\phi_1(\mathbf{r}) + \nu\Sigma_{f2}\phi_2(\mathbf{r}) \right ) \\
 -\nabla D(\mathbf{r})_2 \nabla \phi_2 (\mathbf{r}) + \Sigma_{a2}(\mathbf{r}) \phi_2 (\mathbf{r}) 
    &= \Sigma_{s1\to 2}\phi_1(\mathbf{r}) \, ,
 \end{split}
 \label{eq:twogroupdiff}
\end{equation}
where the notation is standard \cite{duderstadt650nuclear}, and all parameters are defined in the benchmark documentation \cite{center1977benchmark}.
A mesh-centered, finite-volume approximation was used with a uniform, $45 \times 45$ spatial mesh.  
Upon discretization, the entire set of equations was cast in terms of the fission source density, i.e., $\mathbf{f} = \nu\Sigma_{f1}\phi_1(\mathbf{r}) + \nu\Sigma_{f2}\phi_2(\mathbf{r})$, which results in a $2025 \times 2025$ operator. 
Hence, the problem is by no means a large one, but it proved to be a valuable test case for the method.

All calculations were initialized with a starting vector of which each element was drawn from the uniform distribution $U[0, 1]$. 
This randomized starting vector helps to ensure that all eigenmodes can be present.
A formal sensitivity study was not performed to understand how this initial guess impacts the algorithm performance, but scoping studies suggest there is little impact on the numbers of iterations required for any particular algorithm.
In all cases, a reference solution was computed using the implicitly-restarted Arnoldi method as implemented in SciPy \cite{scipy}.
All DMD calculations were performed using {\tt PyDMD} \cite{pydmd}.

\subsection{Skipping Ahead with DMD-PM($n$)}

\begin{figure*}[t!]
 \centering
 \includegraphics[width=0.8\textwidth]{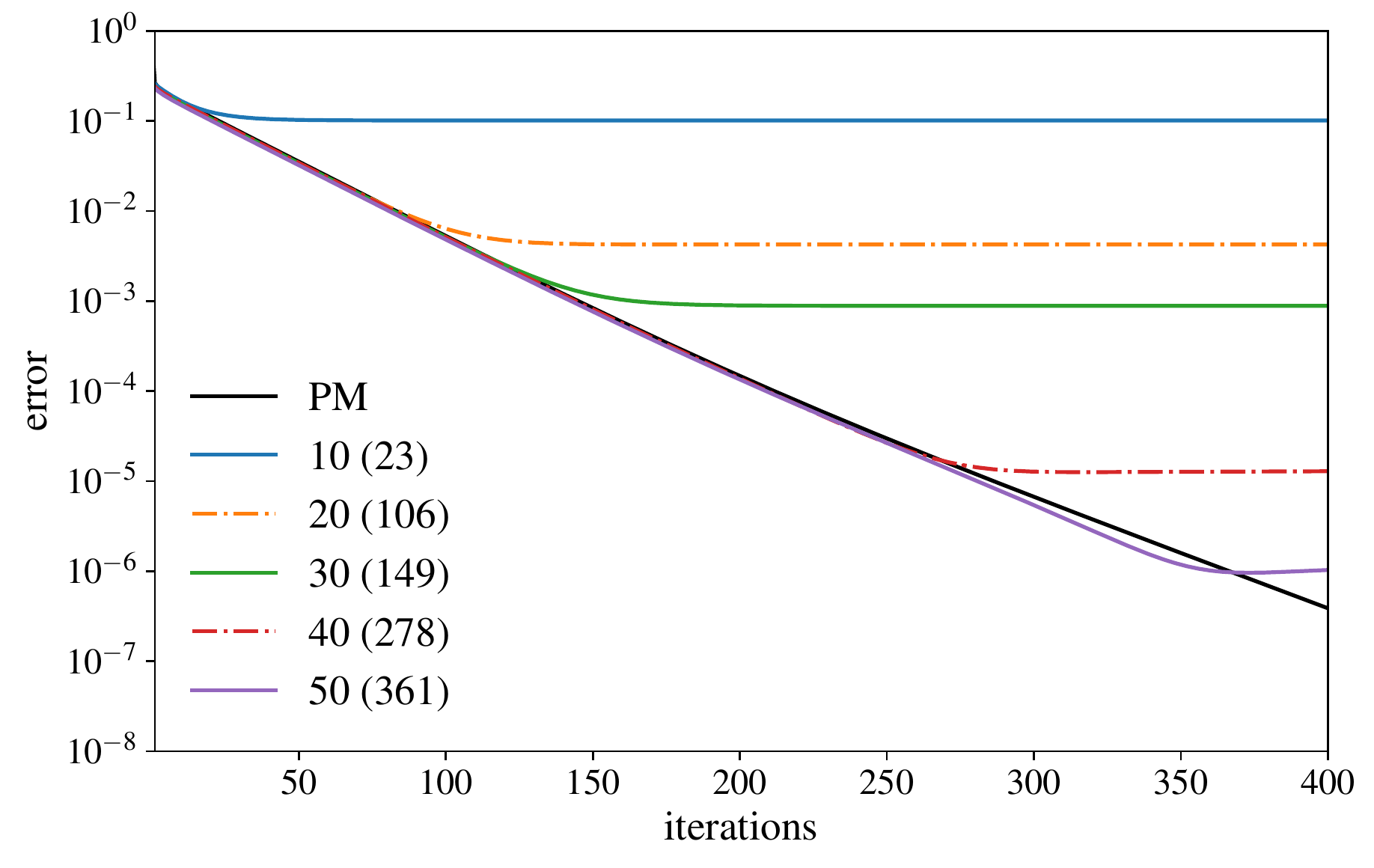}\\
   \caption{Error in the DMD-predicted, dominant eigenmode as a function of iteration.  The legend shows the number of power iterations $n$ used to construct the surrogate, and in parentheses is the effective number of power iterations the DMD surrogate can produce.}
  \label{fig:skipahead}
\end{figure*}

As a first test of the method, a series of $n$ power iterations were performed, from which a DMD surrogate following Eq.~(\ref{eq:dmd_surrogate}) was defined.
The dominant eigenmode was reconstructed as a function of iteration, and the error with respect to the reference eigenmode was computed.
Shown in Figure \ref{fig:skipahead} are the results for several values of $n$.  
Errors in the eigenmode as computed from standard power iterations are also shown. All errors are defined as $||\mathbf{x}^{\text{reference}}-\mathbf{x}^{\text{approximate}}||_2$.

The DMD surrogates reproduce the eigenmodes very accurately for $n$ or fewer iterations.
The error approaches an asymptotic, lower bound as predictions are made beyond the number of power iterations used to generate the DMD surrogate.
Shown in parentheses are the number of equivalent power iterations to which the final, saturated error in the DMD prediction corresponds.
For example, application of 30 power iterations leads to a DMD surrogate that can predict an eigenmode with an accuracy equal to 149 power iterations, a substantial skip ahead in the number of iterations.  
This skipping ahead is precisely what the algorithm described in Section~\ref{sec:dmdpi} facilitates.

\subsection{Application of Restarted DMD-PM$(n)$}

\begin{figure*}[t!]
 \centering
 \includegraphics[width=0.8\textwidth]{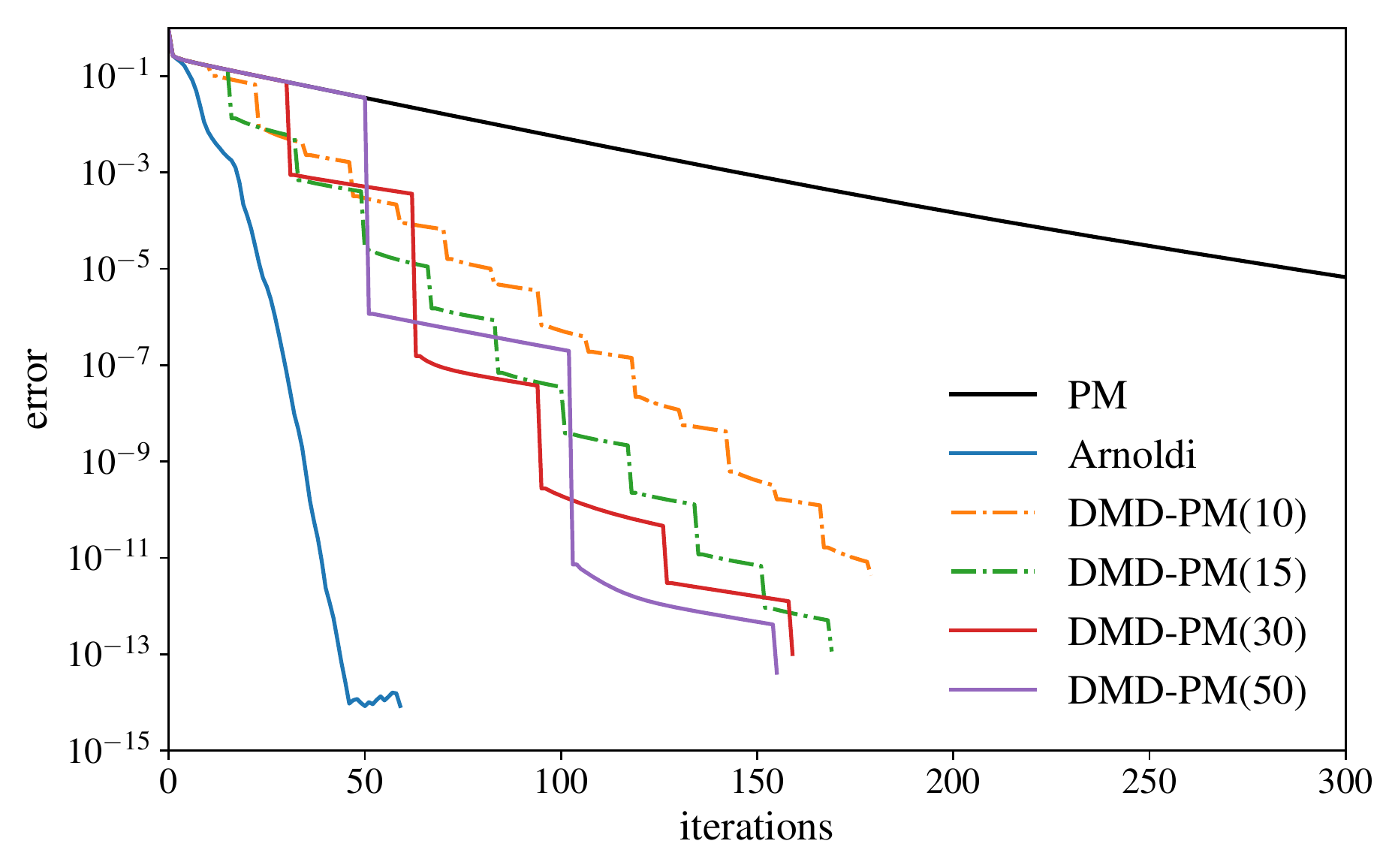}\\
   \caption{The error in the predicted eigenmode for DMD-PM($n$), where $n$ is the number of power iterations performed.  Errors are also included for the power method (PM) and Arnoldi's method.}
  \label{fig:dmdpi_semilog}
\end{figure*}

To test the iterative application of the DMD-accelerated power method, the same problem was solved for a variety of restart values $n$.
The results are shown in Figure~\ref{fig:dmdpi_semilog}.
Also included are errors for the power method (PM) and Arnoldi's method. 
Here, the Arnoldi method was used without restarts.
The results shown for Arnoldi are as a function of the size of the subspace used.

The restart values for DMD-PM($n$) were selected so that the same number of matrix-vector multiplications $\mathbf{A}\mathbf{v}$ were used to construct the underlying DMD surrogate.
Here, the extrapolation is counted as an iteration (along the horizontal axis), so the final end points do not match exactly.
Although all four applications of DMD-PM($n$) significantly accelerate the power method, use of a larger restart value appears to produce slightly better acceleration (at the cost, of course, of storing a larger number of snapshots).
For reference, approximately 800 power iterations are required for similar, final errors (i.e., $10^{-14}$).

A stark difference can be observed between the DMD-PM($n$) performance and that of Arnoldi's method.  Whereas DMD-PM($n$) requires more than 150 iterations to converge to within, say, $10^{-13}$, Arnoldi's method requires fewer than 40.
This difference is not altogether surprising.  
Arnoldi is based on a subspace that undergoes continuous orthonormalization, which produces a better-conditioned and, likely, a richer basis than can be produced by successive application of $\mathbf{A}$ to a single vector.

\subsection{Restarted DMD-PM($n$) for Higher Modes}

Like Arnoldi's method (and others based on construction of Krylov or other subspaces), DMD-PM($n$) can recover higher-order modes, at least approximately.
However, an unrestarted DMD-PM approximation leads to an ill-conditioned basis and, hence, cannot produce approximations for higher-order modes with reliable accuracy.
Moreover, the iterative DMD-PM($n$) is, by design, ill suited for recovering higher-order modes because it essentially throws away all higher-order modes upon the restart.

\begin{figure*}[h]
 \centering
    \includegraphics[width=0.65\textwidth]{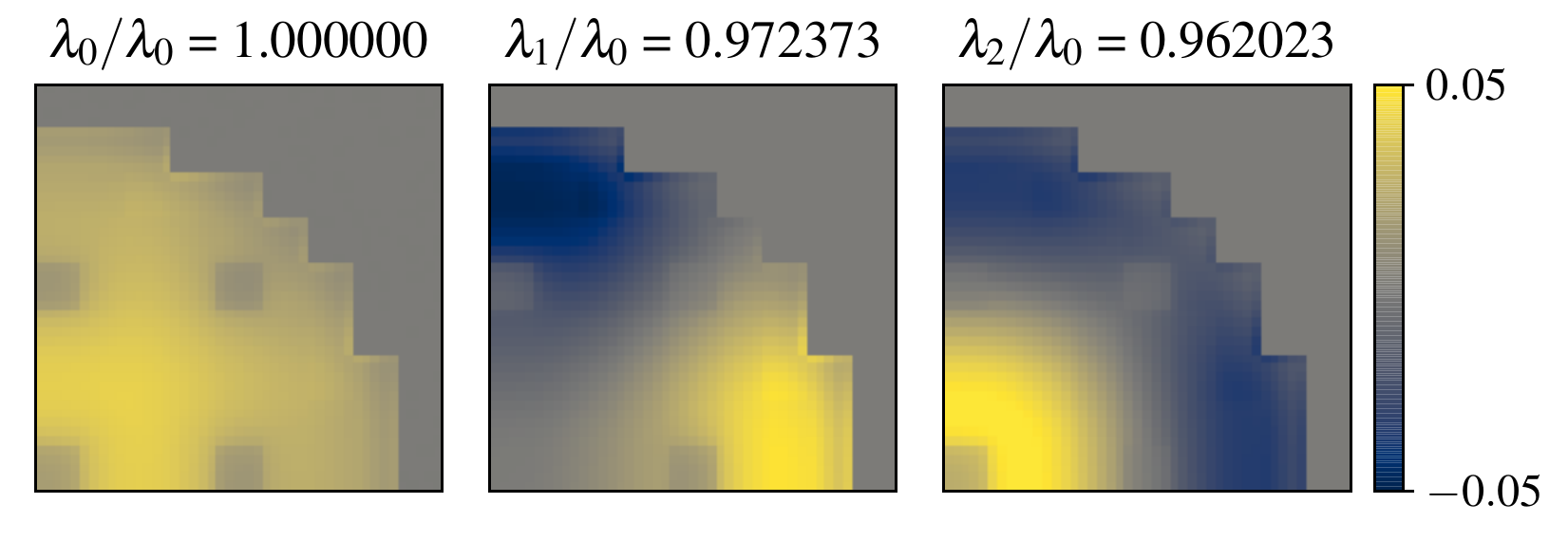}\\
    \caption{First three reference eigenmodes; the corresponding eigenvalue ratios are shown above.}
    \label{fig:ref_modes}
\end{figure*}

\begin{figure*}[!]
 \begin{subfigure}[b]{\textwidth}
  \centering
    \includegraphics[width=0.7\textwidth]{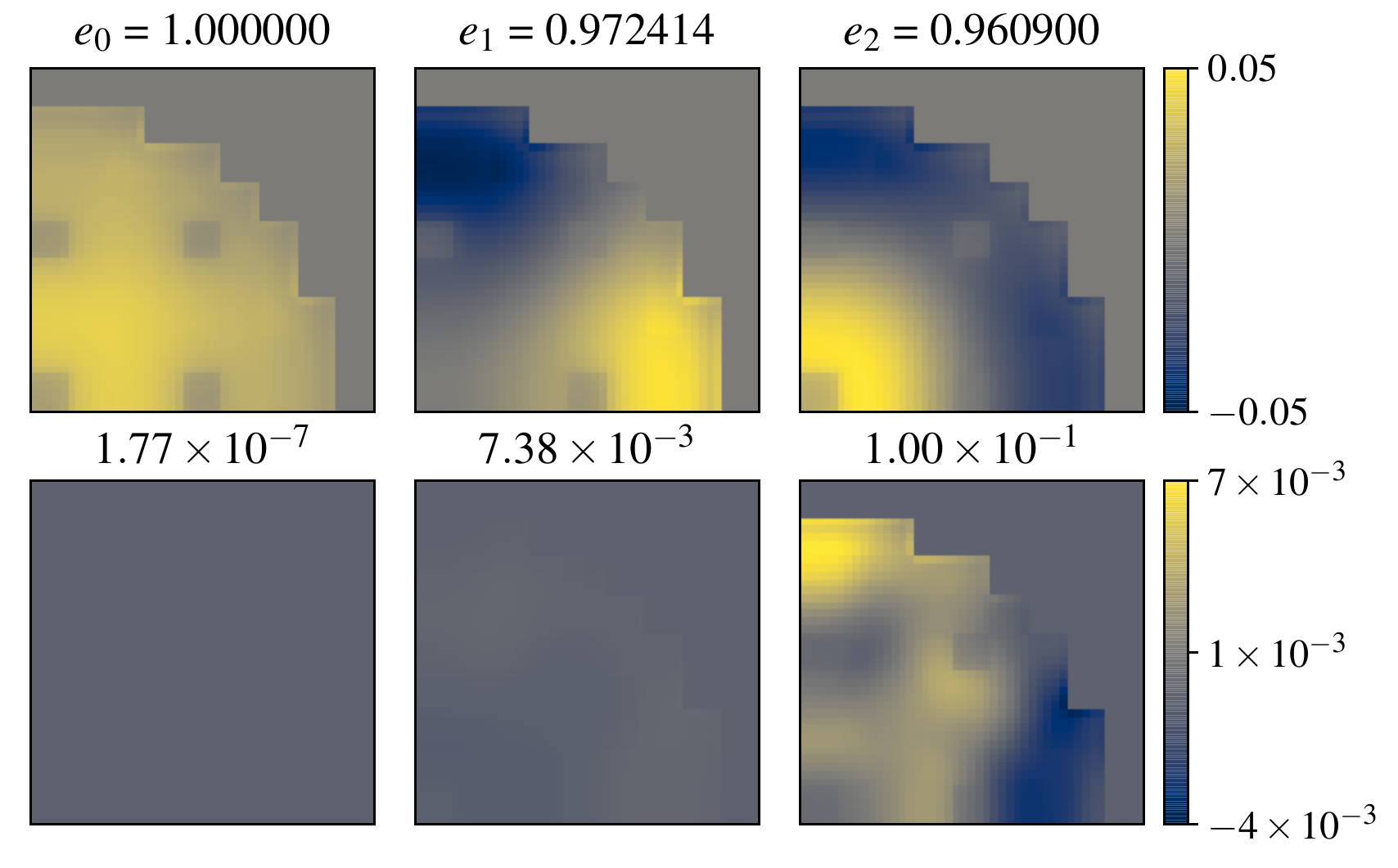}\\
    \caption{Two iterations of DMD-PI(30) with retention of 3 approximate modes.}
    \label{fig:appx_modes_2_3}
 \end{subfigure}
 
 \begin{subfigure}[b]{\textwidth}
  \centering
   \includegraphics[width=0.7\textwidth]{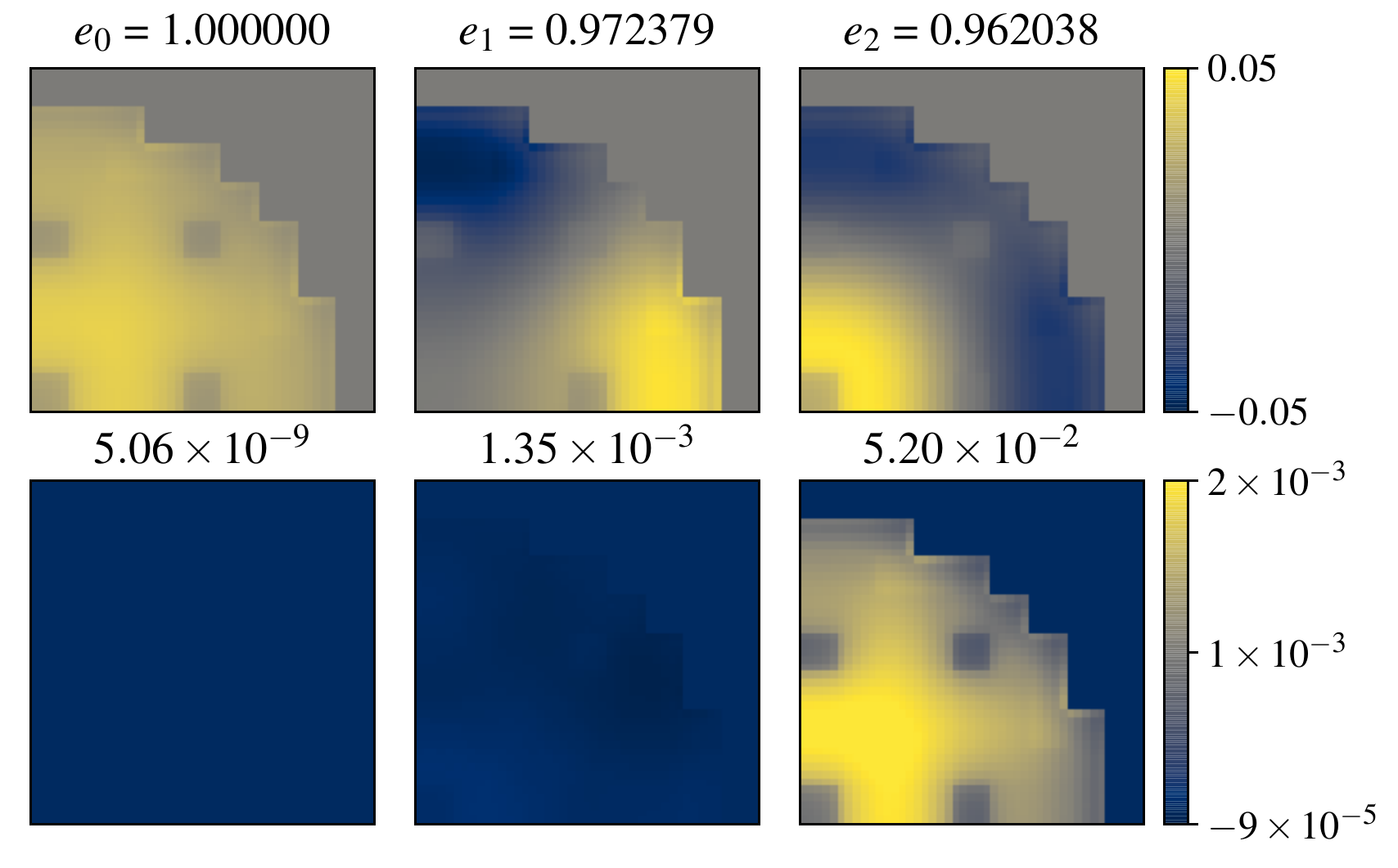}\\
   \caption{Three iterations of DMD-PI(30) with retention of 3 approximate modes.}
   \label{fig:appx_modes_3_3}
 \end{subfigure}
 
 \begin{subfigure}[b]{\textwidth}
  \centering
   \includegraphics[width=0.7\textwidth]{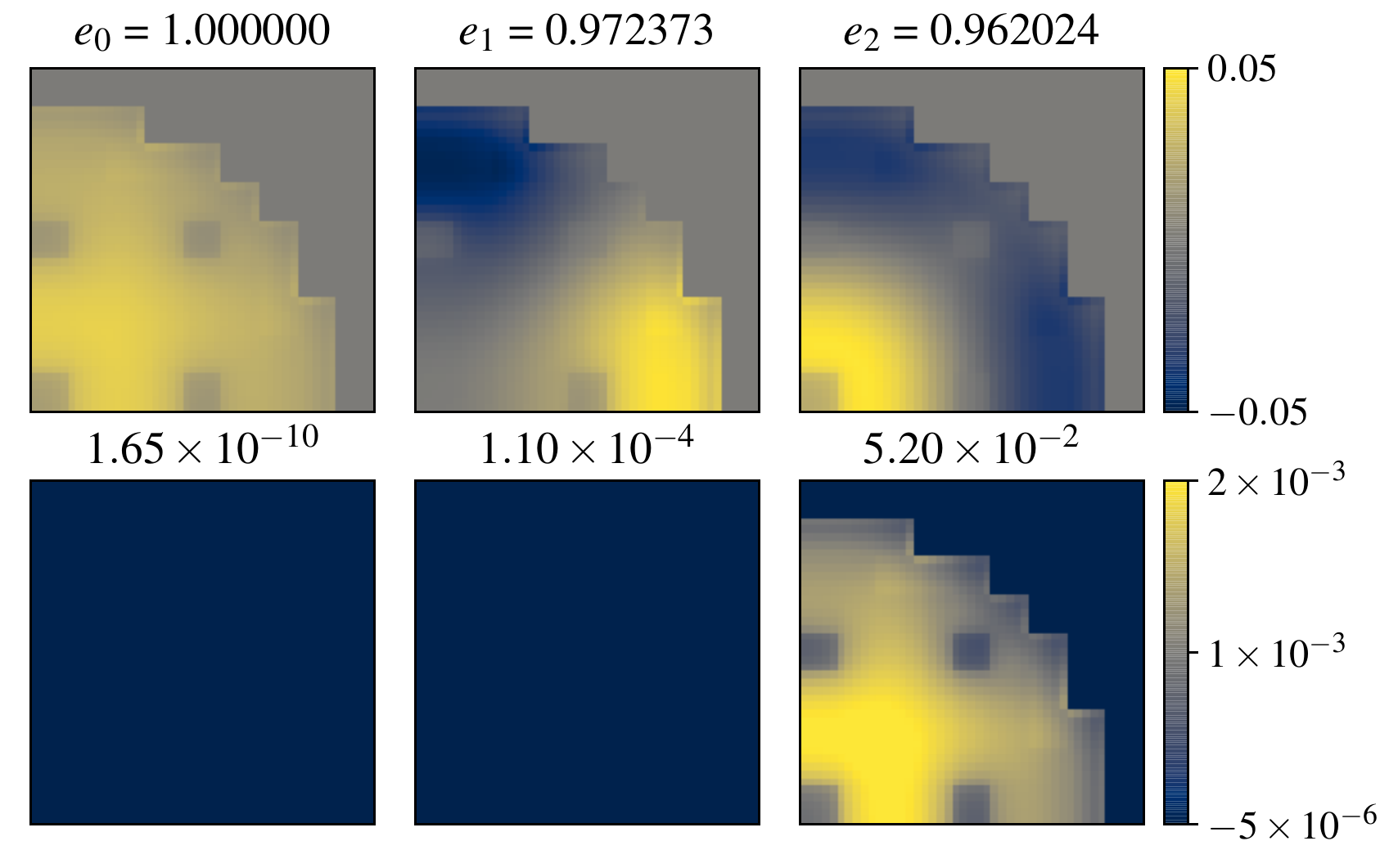}\\
   \caption{Four iterations of DMD-PI(30) with retention of 3 approximate modes.}
   \label{fig:appx_modes_4_3}
 \end{subfigure}
 \caption{Shown in the top row of each subfigure are the first three modes as computed from several applications of DMD-PM($30$).   The second row shows the error $\mathbf{e}_i = \mathbf{x}_i^{\text{reference}}-\mathbf{x}_i^{\text{approximate}}$.  The corresponding $e_i \approx \lambda_i/\lambda_0$ (top row) and norm of the error $||\mathbf{e}_i||_2$ (bottom row) are also shown.}
 \label{fig:appx_modes}
\end{figure*}

Consequently, a slight variation of the iterative, restarted DMD-PM($n$) algorithm was tested.
Rather than keeping only the DMD mode corresponding to the largest eigenvalue, the dominant mode was kept with a small contribution from the next two modes in order to capture the three modes with the largest eigenvalues.
Specifically, the initial guess between restarts was chosen to be the $\mathbf{z}_0 + \epsilon ( \mathbf{z}_1 + \mathbf{z}_2)$, where $\epsilon$ is a small value (here, $10^{-4}$) that guarantees that the next iteration is started with at least some contribution from the higher-order space.
The largest three eigenvalues and their corresponding modes were recovered from the DMD calculation after the second, third, and fourth iterations of DMD-PM($30$) as approximations for the first three eigenpairs of the original system.
Shown in Figure~\ref{fig:ref_modes} are the reference modes.
The corresponding, absolute errors in the DMD-PM($n$) approximations are shown in Figure~\ref{fig:appx_modes}.
All computed eigenvectors were normalized to unity.  
Errors in higher-order mode estimates were found to depend somewhat on the randomized initial guess, but those shown are representative values.

As can be expected, the errors in the two higher-order modes (and their eigenvalues) are much larger than the error for the dominant mode.
Moreover, these errors decrease somewhat with each iteration.
The error in the dominant mode after two iterations ($1.77\times 10^{-7}$) is nearly unchanged from the case in which higher modes are not kept ($1.54\times 10^{-7})$; see Figure \ref{fig:dmdpi_semilog}.
However, the performance does degrade somewhat thereafter, with errors after three and four iterations of approximately $5.06\times 10^{-9}$ and $1.65\times 10^{-10}$, respectively, compared to $9.90\times 10^{-11}$ and $3.00\times 10^{-12}$ in Figure~\ref{fig:dmdpi_semilog}.

\section{Conclusions}

The DMD-PM($n$) method was found to provide reasonable ($5 \times$) speedup compared to unaccelerated power iterations.
Although not competitive with Arnoldi for the test problem studied, there do exist applications for which access to iterates is only available in a post-processing sense. 
In reactor analysis, the use of the power method in Monte Carlo simulations is widespread.  
Application of DMD-PM($n$) in that setting may be an obvious solution for convergence acceleration (a common problem) and, potentially, a tool for analysis of source convergence and tally uncertainties.
Also in that setting, the continual perturbation of higher modes due to uncertainties may make their computation easier and more robust (relative to the deterministic examples shown above).

\section*{Acknowledgments}

The material presented is based on work supported in part by a Faculty Development Grant awarded to Kansas State University by the U.S. Nuclear Regulatory Commission under award number NRC-HQ-60-17-G-0010.

\bibliographystyle{siamplain}
\bibliography{references}
\end{document}